\documentclass[aps,prb,twocolumn,ssecnumarabic]{revtex4-2}
\usepackage{bm}
\usepackage{graphicx}
\usepackage{color}
\usepackage{braket}
\usepackage{amsmath,amssymb,amsfonts,amsthm,mathtools}
\usepackage{enumerate}
\usepackage{enumitem}
\usepackage[colorlinks=true,linkcolor=blue,anchorcolor=red,citecolor=blue,urlcolor=blue]{hyperref}
\usepackage{diagbox}
\usepackage{titlesec}
\usepackage{tikz-cd}
\usepackage{float}
\usepackage{multirow}
\usepackage[title]{appendix}
\usepackage{pdfpages}
\usepackage{changes}
\usepackage{stackengine}
\usepackage{makecell}

\makeatletter
\AtBeginDocument{\let\LS@rot\@undefined}
\makeatother

\makeatletter 
\renewcommand\NAT@biblabelnum[1]{#1.} 
\makeatother

\newcommand{\mr}[1]{\mathrm{#1}}
\newcommand{\mbf}[1]{{\bm #1}}
\newcommand{\mcal}[1]{\mathcal{#1}}

\begin{document}

\title{Spin pumping effect in non-Fermi liquid metals}

\author{Xiao-Tian Zhang$^{1,7}$}\email{zhangxiaotian@ucas.ac.cn }

\author{Yi-Hui Xing$^{2,7}$}

\author{Xu-Ping Yao$^{1}$}

\author{Yuya Ominato$^{3}$}

\author{Long Zhang$^{1,4}$}\email{longzhang@ucas.ac.cn}

\author{Mamoru Matsuo$^{1,5,6}$}\email{mamoru@ucas.ac.cn}
\affiliation{$^{1}$Kavli Institute for Theoretical Sciences, University of Chinese Academy of Sciences, Beijing 100190, China}
\affiliation{$^{2}$Institute of Physics and University of Chinese Academy of Sciences, Chinese Academy of Sciences, Beijing 100190, China}
\affiliation{$^{3}$Waseda Institute for Advanced Study, Waseda University, Shinjuku, Tokyo 169-8050, Japan.}
\affiliation{$^{4}$CAS Center for Excellence in Topological Quantum Computation, University of Chinese Academy of Sciences, Beijing 100190, China}
\affiliation{$^{5}$RIKEN Center for Emergent Matter Science (CEMS), Wako, Saitama 351-0198, Japan
}%
\affiliation{$^{6}$Advanced Science Research Center, Japan Atomic Energy Agency, Tokai, 319-1195, Japan
}%
\affiliation{$^{7}$ These authors contributed equally: Xiao-Tian Zhang and Yi-Hui Xing.}%

\date{\today}

\begin{abstract}
Spin pumping effect is a sensitive and well-established experimental method in two-dimensional (2D) magnetic materials. We propose that spin pumping effect can be a valuable probe for non-Fermi liquid (NFL) behaviors at the 2D interface of magnetic heterostructures. We show that the modulations of ferromagnetic resonance exhibit power-law scalings in frequency and temperature for NFL metals induced near a quantum critical point (QCP). At the Ising nematic QCP, we demonstrate that the enhanced Gilbert damping coefficient $\delta \alpha$ acquires negative power-law exponents in distinct frequency regimes. The exponents convey universal parameters inherited from the QCP and reflect the non-quasiparticle nature of the spin carriers in the NFL metal. At finite temperature, we show that the Gilbert damping mechanism is restored in the quantum critical regime and $\delta \alpha$ measures the temperature dependence of the correlation length. 
Our theoretical proposal has the potential to stimulate the development of an interdisciplinary research domain where insights from non-equilibrium spin physics in spintronics are integrated into strongly correlated matter.

\end{abstract}

\maketitle

\noindent \textbf{\large INTRODUCTION}\label{sec:introduction}\\
Dimensionality plays a vital role in many-body physics, particularly,
the universal critical phenomena near a quantum critical point (QCP)~\cite{Sachdev}.
Two-dimensional (2D) correlated systems with strong quantum fluctuation have been 
a fertile land for various unconventional quantum phases of matter and associated QCPs,
including the cuprate oxide layers in high-$T_{\rm c}$ superconductors~\cite{taillefer2010scattering, keimer2015quantum}, 
non-Fermi liquid metals~\cite{SSLee2018} and exotic quantum magnets~\cite{Burch2018}. 
The atomic-level 2D nature facilitates the formation of the proximity effect,
thus, offering a practical platform for co-integrating distinct 
physical ingredients by artificial design of heterostructures~\cite{Novoselov2016}.
The stacking and twisting of 2D materials have expanded 
the boundary of condensed matter physics 
and give birth to the van der Waals (vdW) heterostructure~\cite{Burch2018} and ``twistronics''~\cite{Andrei2021}.
The lower dimensionality enables otherwise unattainable fabrication, 
manipulation, and measurement as their bulk counterparts~\cite{Hellman2017}.
The strain~\cite{Cenker2022}, gating~\cite{Huang2018}, light~\cite{Sunko2020} and electric field~\cite{Huang2018} can couple with various internal degrees of freedom --- charge, spin, orbit, lattice, etc.
This offers unprecedent tunability towards QCPs.
In the mediated quantum critical region,
wild quantum fluctuations can lead to Fermi liquid instability
for the conduction electrons~\cite{RMP2007}.
The breakdown of coherent Fermi-liquid quasiparticles~\cite{Landau} 
is the most dramatic manifestation of the many-body correlation, 
which is known as non-Fermi liquid (NFL) behavior.

Despite the success in fabricating 2D magnetic thin layers,
the experimental detection of the critical spin fluctuation poses significant challenges.
The conventional magnetic probes, such as neutron scattering, superconducting quantum interference device magnetometry, are designed for bulk magnets,
therefore, are insensitive provided the weak signals from magnetic vdW materials.
The effectiveness and efficiency of optical probes, such as Raman spectroscopy or magneto-circular dichroism, are currently under investigation~\cite{Mak2019}.
It is greatly desired that a clever implementation of the magnetic vdW heterostructure
probes the many-body correlation in directly in the spin channel.
Many-body correlations have been probed using the pure spin current in the spintronics community~\cite{Han2019}.
The interface at the heterostructure of various 2D magnetic materials 
is particularly beneficial for harnessing spintronic functionalities~\cite{Sierra2021}.
The spin current can be mediated by various quasiparticles given the rich combination of heterstructures.

The spin pumping effect~\cite{Tserkovnyak2002, Tserkovnyak2005}, originally designed at the interface of magnetic heterostructure to generate pure spin current, develops potential as a sensitive probe in measuring the magnetic phase transitions in 2D thin layers~\cite{Qiu2016}.
The heterostructure comprises a magnetic insulator subjected to dynamical driven field
and a 2D quantum magnet as the spin current receiver.
Conventionally, one consider a ferromagnetic insulator (FI) with a precessional magnetization at its resonance states, the spin current is injected into the adjacent material
via interfacial spin exchange interaction~\cite{Ohnuma2014, Matsuo2018, Kato2019}.
The spin injection has a backaction on the FI in modulating the frequency of feromagnetic resonance (FMR) and the Gilbert damping. 
The modulated FMR signal carries information about the dynamical spin susceptibility 
of the 2D magnetic thin layer, making it a effective probe for studying its spin characteristics of 2D magnetic materials.
It is therefore appealing to probe the spin correlation, particularly the critical spin fluctuation near the QCP, in 2D magnetic heterostructure using the spin pumping.

\begin{figure}[htbp]
	\centering
	\includegraphics[width=\linewidth]{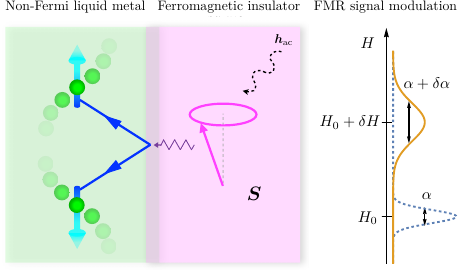}
	\caption{Feromagnetic resonance-driven spin pumping effect in the magnetic heterostructure considered in this work. 
	(a) Schematic plot of the NFL/FI bilayer structure for the FMR-driven spin pumping experiment. The pink arrow in the FI indicates the spin ${\bm S}$ precessing in the external ac magnetic field ${\bm h}_{\rm ac}$. The blue arrows in the NFL metal indicate the itinerant electrons exchanging spin angular momentum at the interface with magnons in the FI layer. The gradually fainted green balls illustrate the incoherent quasiparticles in the NFL metal.
    (b) The FMR signal modulation due to the interlayer coupling, where $H$ is the magnitude of the Zeeman field and $\alpha$ is the coefficient of the Gilbert damping.
    }%
	\label{fig:illustration}
\end{figure}

In this work, we consider the spin pumping effect in the magnetic heterostructure composed of a 2D NFL metal and a ferromagnetic insulator (FI) thin film schematically plotted in Fig.~\ref{fig:illustration}.
We demonstrate that the spin pumping is an effective method to probe the critical 
spin fluctuation for 2D magnetic heterostructure in the dynamic regime near the QCP.
We consider the type of QCP near a Pomeranchuk instability
which is described by gapless Fermi surface coupled with critical bosons.
We begin with generic analysis based on universal scaling rule near the QCP;
Then, we take the 2D Ising nematic QCP induced NFL as a concrete example
which is relevant to underdoped cuprates~\cite{keimer2015quantum} and Fe-based superconductors~\cite{Si2016}. 
The FI is at the resonant frequency under the microwave radiation,
the magnon excitations are damped by the phenomenlogical Gilbert mechanism. 
Pure spin current is injected into the adjacent NFL
where the spin angular momentum is carried by ``non-quasiparticles''.
The backaction of the spin injection can be read off from the magnon self-energy correction.
The perturbation is carried out up to second order in interfacial exchange coupling in the Keldysh representation
which gives rise to a direct relation between the magnon self-energy and the dynamical spin susceptibility in the NFL metal.

We evaluate the dynamical spin susceptibility $\chi_{\rm uni}(\omega)$ near Ising nematic QCP
when the itinerant spins in NFL metal are subjected to the interfacial exchange interaction from the FI.
The interfacial spin exchange interaction not only facilitates the spin injection and FMR modulations
but also provides a relaxation channel for the itinerant spins.
Thus, it promotes the spin dynamics in the uniform component
which conceives the non-quasiparticle nature of the underlying NFL metal.
To this end, we deal with the fermion-boson coupling near the QCP in a self-consistent manner,
and then we obtain power-law scalings for $\chi_{\rm uni}(\omega)$ perturbatively in terms of interfacial exchange interaction.
The FMR signals, namely the shift of resonance frequency and enhanced Gilbert damping coefficient, are significantly modulated:
The enhanced Gilbert damping coefficient exhibits a power-law divergence in the low-energy limit
indicating that the conventional Gilbert mechanism becomes invalid.
The scaling exponents are intimately related to the university class of the underlying QCP.
At finite temperatures, we adopt revised Eliashberg equations for fermions and bosons
where bosonic correlation length and electron scattering rate acquire characteristic temperature dependence.
In the quantum critical regime, the Gilbert damping is restored and the temperature dependence
in the enhanced Gilbert damping coefficient captures the distance towards the QCP.
We show out that, at both zero and finite temperatures, 
the shift of resonance frequency captures characteristics of quasi-particle disappearance.

\bigskip
\noindent \textbf{\large RESULTS}\label{sec:results}\\
\textbf{Magnetic Heterostructure}.\label{subsec_magnetic_heterostructure}
Let's consider the magnetic heterostructure composed of a NFL metal 
and a FI thin film shown in Fig.~\ref{fig:illustration}(a). 
The FI is driven by an external \emph{ac} magnetic field in resonance with the precession of local spins. 
The spin current is injected into the NFL metal via the spin exchange coupling at the interface. 
The full Hamiltonian of the NFL/FI heterostructure comprises three parts,
\begin{equation}\label{Ham}
    {\cal H}(t) = {\cal H}_{\rm FI}(t) + {\cal H}_{\rm ex}+ {\cal H}_{\rm NFL}.
\end{equation}
The first term describes the FI with the ferromagnetic Heisenberg model subjected to an oscillating magnetic field,
\begin{multline}
    {\cal H}_{\rm FI}(t) = - J \sum_{\langle i,j\rangle} {\bm S}_i \cdot {\bm S}_j + \gamma_{\rm g} H \sum_{i}S_i^z \\
        - \gamma h_{\rm ac}\sum_i \big( S_i^x  \cos \omega t - S_i^y \sin \omega t\big),
\end{multline}
in which $J>0$ is the ferromagnetic exchange coupling constant. ${\bm S}_i$ stands for the local spin at site $i$ in the FI. $H$ is the magnitude of the Zeeman field. $\gamma_{\rm g}$  ($<0$) is the gyromagnetic ratio. $h_{\rm ac}$ and $\omega$ are the amplitude and the frequency of the circularly oscillating external magnetic field, respectively. The second term in Eq.\eqref{Ham} is the exchange coupling at the interface between local spins in the FI and itinerant electrons in the NFL metal,
\begin{align}
    {\cal H}_{\rm ex} &= \sum_i \int d^2{\bm r} \ J({\bm r},{\bm r}_i) {\bm S}_i\cdot {\bm s}({\bm r}) \notag \\
    & = \sum_{{\bm k},{\bm q}} \big( J_{{\bm k},{\bm q}}  s_{\bm q}^+ S_{\bm k}^- + {\rm H.c.} \big) + \cdots, 
    \label{H_ex}
\end{align}
where ${\bm s}({\bm r}) = \frac{1}{2}c^\dagger_{\alpha}({\bm r}) \mbf{\sigma}_{\alpha \beta} c_{\beta}({\bm r})$ is the itinerant electron spin operator. 
The exchange coupling function in the reciprocal space is approximately given by~\cite{Ominato2022pSC}
\begin{equation}\label{J_J1}
    |J_{{\bm k},{\bm q}}|^2= \frac{J_1^2}{N} \delta_{{\bm k},{\bm q}} + \frac{J_2^2 l^2}{AN},
\end{equation}
where $A$ is the area of the NFL at the 2D interface, 
and $N$ is the number of sites in FI. 
The first and second terms describe averaged uniform 
and spatially uncorrelated roughness respectively.
$l$ is introduced as an atomic scale length for the continuous model.
$J_{1}$ and $J_2$ are the mean value and the variance of the exchange coupling.

The third term in Eq.(\ref{Ham}) is a fermion-boson coupled model
describing the NFL induced by critical boson modes.
The mechanism and classification of different types of NFL 
has a long and distinguished history~\cite{Hertz1976,Millis1993,Polchinski1994,SSLee2018}
with ever-increasing new developments~\cite{Sur2019,Lake2021,Zhang2023,Zhang2023B,PAN2024116451}.
The critical boson induced NFL can be described by a generic form of Hamiltonian,
\begin{multline}\label{Ham_NFL}
    \mcal{H}_{\rm NFL} = \int \mathrm{d}^2{\bm r} \Big[ c^\dagger_\alpha({\bm r}) \epsilon(-i\partial_{{\bm r}})c_\alpha({\bm r}) - \lambda O({\bm r})\phi({\bm r}) \\
    + \frac{1}{2} \big(\partial_{\bm r} \phi\big)^2 + \frac{r}{2} \phi^2 + \cdots \Big],
\end{multline}
Here $c^\dagger_\alpha$($c_\alpha$) is the electron creation (annihilation) operator for spin the spin up/down $\alpha=\pm$. $\epsilon({\bm k})$ is the spin degenerated, bare electronic dispersion near the Fermi surface.
$\phi$ is a critical fluctuating bosonic field admitting a Ginzburg-Landau expansion
near a QCP tuned by $r\rightarrow 0$.
$\lambda$ is the coupling constant of electrons and the bosonic order parameter field. 
$O({\bm r})$ is a fermion bilinear operator, which transforms inversely as $\phi(\mbf{r})$ 
under symmetry actions guaranteeing the invariance of the coupling term.
We note that the boson field $\phi({\bm r})$ is \emph{not} directly coupled to the FI via magnetic exchange interactions.
It's reasonable to assume that the boson field is either regarded as non-magnetic for the Ising nematic case, 
or being ineffective in coupling to the local magnetic moment ${\bm S}_i$ in FI.
Invoking the direct coupling of both itinerant spins and local moments, {\sl e.g.} near the magnetic QCPs,
would introduce additional complexity, which is left for future studies.

\begin{figure}[tb]
	\centering
	\includegraphics[width=\linewidth]{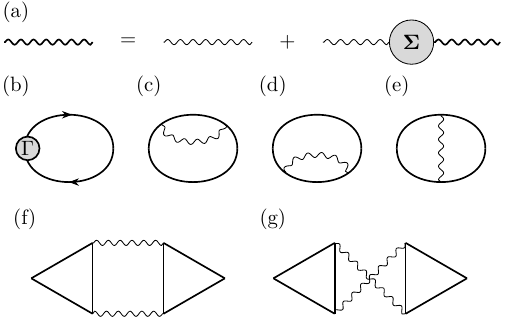}
	\caption{Feynman diagrams for perturbative calculations. (a) Dyson equation for the magnon Green's function. The thin and the thick wavy lines are the bare and the renormalized magnon Green's functions, respectively. The shaded circle represents the magnon self-energy $\Pi$.
    (b) The fully renormalized charge polarization bubble. Thick straight lines are the fermion propagators and shaded circle represents the vertex function in the fermion-boson model Eq.(\ref{Ham_NFL}).
    (c-g) Feynman diagrams for the charge polarization bubble. At leading order in interfacial exchange coupling $\tilde{J}$,
    there are (c,d) the self-energy diagrams and (e) the vertex diagrams. On the next leading order, the potentially relevant diagrams are (f,g) the Aslamazov-Larkin diagrams.
    }%
	\label{fig:FeynmanDiagramFI}
\end{figure}

\smallskip
\noindent\textbf{Spin pumping and FMR modulations}.\label{subsec_spin_pumping_and_FMT_modulation}
The spin pumping, at a microscopic level, is initiated by the magnetic exchange interaction at the heterostructure interface between the magnetization in the FI and the electron spin of the NFL. This interaction promotes spin injection and generates a self-energy correction for the magnons in the FI as a backaction, modulating the frequency of FMR and Gilbert damping~\cite{Ohnuma2014, Matsuo2018, Kato2019}. 
The modulated FMR signal carries information about the dynamical spin susceptibility 
and is shown to be related to the magnon self-energy.
The Dyson equation for the magnon Green function, illustrated in Fig.~\ref{fig:FeynmanDiagramFI}(a),
is written as
\begin{equation}
\begin{aligned}
    & G^{-1}({\bm k},\omega) = G^{-1}_0({\bm k},\omega) - \Pi({\bm k},\omega), \\
    & G^{-1}_0({\bm k},\omega)  = \omega - \omega_{\bm k} + i\alpha \omega,\\
\end{aligned}
    \label{FI_G}
\end{equation}
where the bare magnon Green function $G^{-1}_0$ has a dispersion that reads $\omega_{\bm k} = Dk^2 - \gamma_{\rm g} H \ (\gamma_{\rm g}<0)$ and the imaginary term proportional to $\omega$~\cite{Abrikosov1975} is known as Gilbert damping with a coefficient $\alpha$~\cite{Tserkovnyak2002}. In addition, there emerges a magnon self-energy $\Pi({\bm k},\omega)$ due to the backaction of the spin injection. The FMR modulation shown in Fig.~\ref{fig:illustration}(b) is determined by the uniform component ($\mbf{k}=0$) of the magnon Green's function, in which the pole dictates the resonance condition, $\omega + \gamma_{\rm g} H - {\rm Re}\Pi_{{\bm k}=0}(\omega) =0$, thus the resonance frequency is shifted by $\delta H=\gamma_{\rm g}^{-1}{\rm Re}\Pi_{{\bm k}=0}(\omega)$. The imaginary part of the self-energy leads to an enhanced Gilbert damping coefficient, $\delta \alpha = -\omega^{-1}\mathrm{Im}\Pi_{{\bm k}=0}(\omega)$.

The magnon self-energy can be calculated perturbatively up to the second order in terms of the external oscillating magnetic field $h_{\rm ac}$ and the exchange coupling $|J_{{\bm k},{\bm q}}|$ [see Supplementary Note 2 for derivations],
\begin{equation}
    \Pi({\bm k},\omega) = -\sum_{{\bm q}} |J_{{\bm k},{\bm q}}|^2 \chi({\bm q},\omega),
\end{equation}
with $\chi({\bm q},\omega) \equiv i \int dt e^{i(\omega+i0^+)t} \Theta(t) \langle [s_{\bm q}^+(t), s_{-{\bm q}}^-(0)] \rangle$
being the retarded spin susceptibility for NFL metals. 
Inserting the exchange coupling function in Eq.\eqref{J_J1}, the magnon self-energy is given by~\cite{Ominato2022pSC}
\begin{equation}\label{Sigma_mag}
    \Pi_{{\bm k}=0}(\omega)= -\frac{J_1^2}{N}\chi_{\rm uni}(\omega) - \frac{J_2^2 l^2}{AN}\chi_{\rm loc}(\omega).
\end{equation}
where two terms corresponds to uniform and uncorrelated roughness contribution of the interfacial exchange interaction. $J_1, J_2$ are the mean value and variance, singling out the uniform and the local components of the dynamical spin susceptibility. They are expressed as $\chi_{\rm uni}(\omega)\equiv \chi({\bm q}=0,\omega) $ and $\chi_{\rm loc}(\omega)\equiv\sum_{\bm q} \chi({\bm q},\omega)$.

In the vicinity of the QCP at $r= r_{\rm c}$, the dynamical spin susceptibility takes the following universal scaling form at $T=0$, $\chi({\bm q},\omega,r-r_{\rm c}) = \xi^{d_{\chi}} {\chi}({\bm q}\xi,\omega\xi^z)$, in which $\xi$ is the spatial correlation length, $d_{\chi}$ is the scaling dimension of the spin susceptibility, and $z$ is the dynamical exponent. At the QCP, the correlation length diverges, leading to the reduced scaling form, $\chi({\bm q},\omega,T=0) = \omega^{-d_{\chi}/z}\tilde{\chi}({\bm q}/\omega^{1/z})$. 
The magnon self-energy correction has different power-law scaling forms
inherited from the uniform and local components in respective limits,
\begin{equation}\label{scale_0}
    \Pi_{{\bm k}=0}(\omega) =
    \begin{cases}
        -\frac{J_1^2}{N} \chi_{\rm uni}(\omega) \simeq \omega^{-d_\chi/z}, & \eta \gg1\\
        -\frac{J_2^2l^2}{AN} \chi_{\rm loc}(\omega) \simeq \omega^{(d-d_\chi)/z}, & \eta \ll 1.
    \end{cases}
\end{equation}
Here $d_{\chi}$ and $z$ are two independent critical exponents which can be uniquely  determined by tuning the interfacial roughness $\eta= J_1\sqrt{A}/(J_2 l)$.
And, $d_\chi$ is often related to the physical dimension $d$
where deviation are encountered when hyperscaling is violated approaching certain QCPs~\cite{Patel2015,Eberlein2016}.
The power-law scalings are in sharp contrast to the conventional linear-in-frequency Gilbert damping term, thus reflects the NFL behavior associated with the QCP. 
In the finite temperature regime mediated by the QCP, a different set of critical exponents $d_{\chi}^{\prime}$ and $z'$ emerge $\chi({\bm q},\omega,r-r_{\rm c},T) = L_\tau^{d_\chi^\prime/z^\prime} \tilde{\chi}\big({\bm q}L_\tau^{1/z^\prime}, \omega L_\tau, \xi(T)/L_\tau^{1/z^\prime}\big)$. $L_{\tau}=1/(k_{\rm B} T)$ is the characteristic scale in the imaginary time. Note that the correlation length (or the boson mass) is originally a quantum parameter, now, acquires a $T$-dependence endowed by higher order magnon interactions.
The uniform and local components are expressed in terms of correlation length as
\begin{subequations}\label{scale_1}
    \begin{align}
        \chi_{\rm uni}(\omega,T) = {}&  \xi(T)^{d_\chi^\prime} \tilde{\chi}\big(\omega/T, \xi(T)T^{1/z'}\big) , \\  
        \chi_{\rm loc}(\omega,T) = {} &  \xi(T)^{(d_\chi^\prime-d)} \tilde{\chi}\big(\omega/T, \xi(T)T^{1/z'}\big).
        \end{align}
\end{subequations}
Henceforth, we consider the smooth interface $\eta \gg1$
and the interfacial exchange interaction is reduced to 
$ {\cal H}_{\rm ex} = J_1/\sqrt{N} \sum_{\bm k} {\bm s}_{\bm k}\cdot {\bm S}_{\bm k}$.
Our focus is the uniform component of the spin susceptibility $\chi_{\rm uni}(\omega)$
that dominates the magnon self-energy correction in Eq.(\ref{Sigma_mag}).
However, for a decoupled Fermi liquid/NFL metal layer with a single source of magnetism,
$\chi_{\rm uni}(\omega)$ vanishes as a consequence of total spin conservation. 
In contrast, the scenario illustrated in Fig.~\ref{fig:illustration}(a) is fundamentally different.
The magnetic heterostructure constitutes a correlated system with two magnetic degrees of freedom.
The interface provides a relaxation channel for itinerant spins
enabling the spin dynamics in the NFL metal.
Additionally, we note that the local component of the spin susceptibility $\chi_{\rm loc}(\omega)$,
at a rough interface is not subject to the strict constraints of spin conservation, 
which is an interesting topic but beyond the scope of this study.

\begin{figure}[b]
	\centering
	\includegraphics[width=0.8\columnwidth]{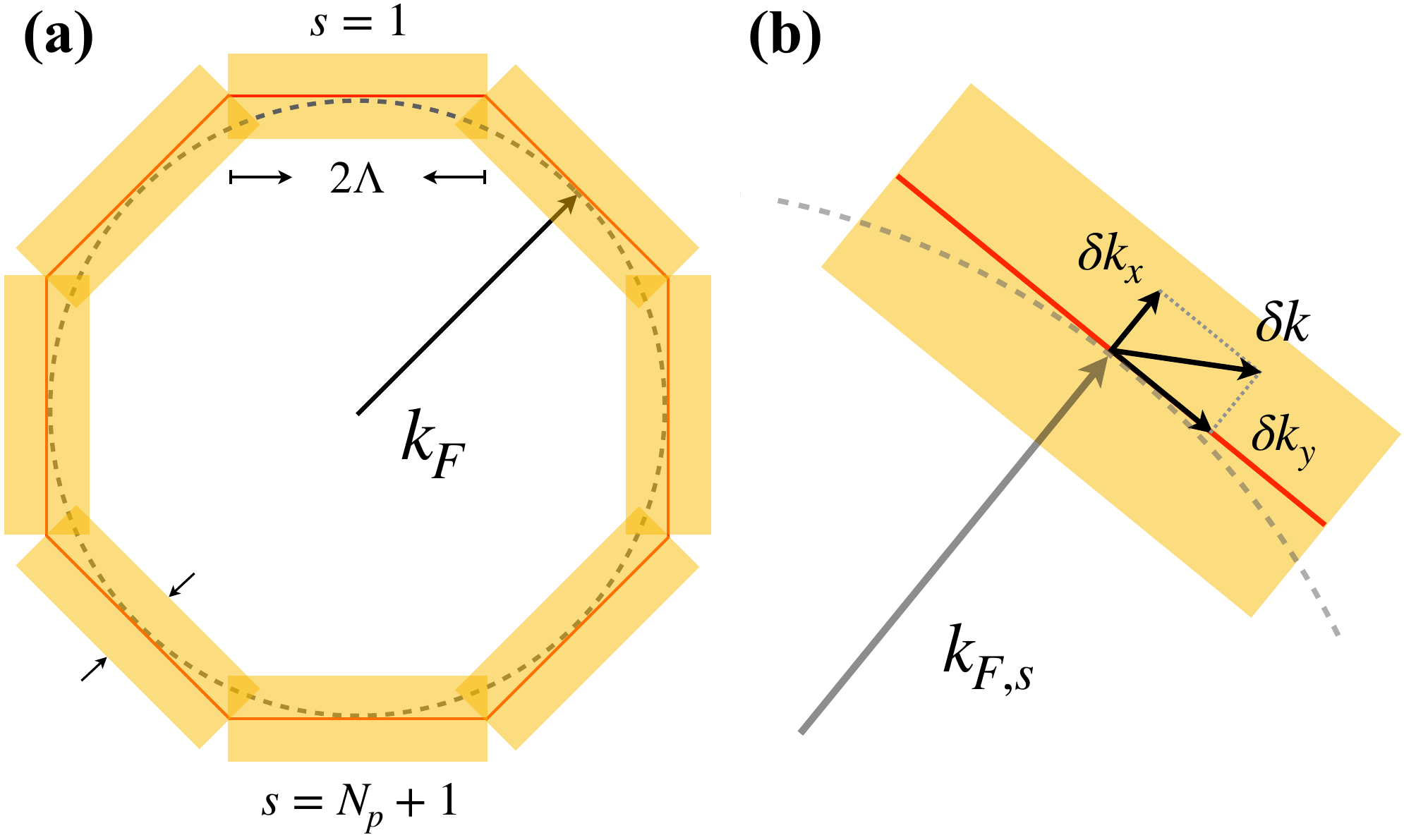}
	\caption{
	Schematic illustration of the patch decomposition
	for 2D circular Fermi surface.
	(a) The critical Fermi circle in 2D is approximated by the orange polygon,
	the dashed grey circle represents the Fermi surface
	and each red line segment stands for decoupled patches with the width $2\Lambda$.
	Orange fainted rectangle is the low-energy region
	for a single decoupled patch labelled by index $s\in[1,2N_{\rm p}]$.
	(b) The small momentum variation $\delta {\bm k}$ in the $s$'th patch centered at ${\bm k}_{{\rm F},s}$ 
	is decomposed into radial component $\delta k_x$ and tangential component $\delta k_y$.
	}
	\label{fig:patch}
\end{figure}

\smallskip
\noindent\textbf{Non-Fermi liquid at the Ising nematic QCP}.
\label{subsec_2}
The Pomeranchuk instability near the 2D Ising nematic QCP
is described by critical bosonic modes of the collectively distorting Fermi surface.
The four-fold lattice rotational symmetry of the 2D Fermi surface 
is on the verge of spontaneously breaking down to two-fold.
The order parameter field $\phi(\mbf{r})$, in the NFL Hamiltonian in Eq.\eqref{Ham_NFL},
is coupled to the fermionic bilinear $O(\mbf{r})$ as
\begin{equation}
    O({\bm r}) = \frac{1}{A}\sum_{\bm q} \sum_{{\bm k},\alpha} d_{\bm k}
    c^\dagger_{{\bm k}-{\bm q}/2,\alpha}c_{{\bm k}+{\bm q}/2,\alpha} e^{i{\bm q}\cdot {\bm r}},
    \label{O}
\end{equation}
where the $d$-wave form factor can take a form $d_{\bm k}=\cos k_x - \cos k_y$.
We adopt the patch decomposition scheme~\cite{SSLee2018}
and divide the Fermi surface into $2N_{\rm p}$ number of patches labelled by $s$-index in Fig.~\ref{fig:patch}(a).
The $s$'th patch is centered at the Fermi momentum ${\bm k}_{{\rm F},s}$
and the momentum nearby is expanded as ${\bm k} = {\bm k}_{{\rm F},s} + \delta{\bm k}$ in Fig.~\ref{fig:patch}(b).
The small variation $\delta \mbf{k}$ can be further decomposed into radial and tangential components $\delta k_{x}, \delta k_{y}$ with respect to the directional vector $\hat{k}={\bm k}_{{\rm F},s}/k_{{\rm F},s}$.
The coupled fermion-boson model specified by Eq.(\ref{Ham_NFL},\ref{O}) admits a self-consistent solution[see Supplementary Note 1 for a self-contained review], where the renormalized fermionic Green function reads 
$g_{s,{\rm R}}(\delta{\bm k},\omega) = \big(ic_{\rm F} d^2_{{\bm k}_{{\rm F},s}} |\omega|^{a} - v_{\rm F} \delta k_x - \frac{\kappa}{2} \delta k_y^2\big)^{-1}$\cite{Metlitski2010} where the spin index $\alpha$ is omitted and the exponent is expressed as $a=\frac{d}{z}=\frac{2}{3}$. 
The dynamical exponent $z=3$ holds up to three-loop perturbations~\cite{Metlitski2010}.
The bare dynamic term $i\omega$ is overwritten by the fermionic self-energy 
$\Sigma_{\rm F}(\mbf{k}_{{\rm F},s},\omega)$ in the low-energy regime $|\omega| < \omega_{\rm c}= c_{\rm F}^3$. 
The real part of the self-energy can be derived using the Kramer-Kronig relation in the low-energy regime as
$ {\rm Re}\Sigma_{\rm F}({\bm k}_{{\rm F},s},\omega) = -\sqrt{3} d^2_{{\bm k}_{{\rm F},s}} c_{\rm F} {\rm sgn}(\omega) |\omega|^{2/3}$.
The NFL feature with $k_{{\rm F},x}\neq \pm k_{{\rm F},y}$ manifests as a vanishing quasiparticle weight:  
$Z_{{\bm k}_{{\rm F},s}}(\omega) \equiv \big[ 1- \partial_\omega {\rm Re} \Sigma_{\rm F}({\bm k}_{{\rm F},s},\omega) \big]^{-1}\simeq d_{\mbf{k}_{{\rm F},s}}^{-2}\omega^{1/3} \xrightarrow{\omega\rightarrow 0} 0$.

The itinerant spins in the NFL are degenerate and 
are \emph{not} on the verge of forming any magnetization
since the critical fluctuation is in the charge channel.
The spin susceptibility is proportional to the charge polarization bubble, 
plotted in Fig.~\ref{fig:FeynmanDiagramFI}(b), 
is given by $\chi_{\mr{uni}}(iq_0) =  -A\sum_{s=1}^{2N_{\rm p}} \Pi_s(iq_0)$.
The charge in each patch is separately conserved provided that the inter-patch scattering is ignored.
A direct evaluation of polarization bubble at $s$'th patch $\Pi_s(iq_0)$
using the fully renormalized fermionic Green function yields a non-vanishing result [see Supplementary Note 2a]
\begin{equation}
\begin{aligned}
	& \Pi_s(iq_0) =  \int \frac{d^2 {\bm k}}{(2\pi)^2} \frac{dk_0}{2\pi}g_{s}( {\bm k},ik_0)  g_{s}( {\bm k},ik_0+iq_0), \\
	& \quad \simeq  - {\rm sgn}(q_0) |q_0|^{-(d-z)/z},\\
\end{aligned}
\label{chi_0}
\end{equation}
where we have kept a shortened notation for the small variation denoted as ${\bm k}$.
However, this naive result contradicts the generic constraint imposed by the spin conservation: 
$\chi_{\mr{uni}}(iq_0)=0$ for any given frequency.
The discrepancy arises from inappropriate and incomplete accounting of the critical boson fluctuations.
By using the renormalized Green function, we take into account of the self-energy corrections in Eq.(\ref{chi_0});
Whereas, the vertex corrections are overlooked, 
which is crucial in maintaining the Ward-Takahashi identity in the dynamic limit $|\Omega|>v_{\rm F}q$.
In fact, at the QCP, if we adopt the renormalized fermion Green function,
the vertex corrections are of order one at any perturbative order in $\mathcal{O}(\lambda)$\cite{Chubukov2005,Chubukov2018}.
Summing up infinite ladder series of vertex corrections in Fig.~\ref{fig:FeynmanDiagramFI}(b) 
yields the correct result for uniform spin susceptibility at Ising nematic QCP, 
\begin{equation}
\chi_{\mr{uni}}(iq_0) =0.
\end{equation}
We derive the cancellation between self-energy 
and vertex in Supplementary Note 2 
where we extend the derivation to a noncircular 2D Fermi surface. 

Furthermore, we turn on the coupling between itinerant spins and local moments 
at the magnetic heterostructure interface [see Eq.(\ref{H_ex})].
The magnetic excitation in the FI is a Goldstone mode of the FM order,
which has been proven ineffective in coupling to quasiparticles in the Fermi liquids\cite{Watanabe2014}.
The NFL phase already formed at Ising nematic QCP is not sabotaged.
As a result, the primary effect of the magnetic excitations 
is the promotion of the spin dynamics in the NFL metal.
To this end, we use the fully renormalized NFL Green function at Ising nematic QCP,
and treat interfacial spin exchange perturbatively.
Since the restriction of spin conservation for the NFL metal is lifted,
the magnetic heterostructure, originally designed for the spin pumping,
represents a natural setup to directly probe the spin dynamics of non-quasiparticles at a Pomeranchuk instability.

\smallskip
\noindent\textbf{Divergent FMR modulation at $T= 0$}.
\label{subsec_3}
The analysis of the spin dynamics at the magnetic heterostructure interface
involves two magnetic degrees of freedom.
The evaluation of spin-polarization bubble for the itinerant electrons perturbatively
in interfacial spin exchange interaction requires renormalizations from
all diagrams depicted in Fig.~\ref{fig:FeynmanDiagramFI}, that includes (a) the self-energy(SE) diagram, 
(b) the vertex(V) diagram and (c) Aslamazov-Larkin(AL)-type diagrams.

To set the stage, let's first consider the case for a free-standing 2D layer of itinerant electrons  
where the magnetism has a single source.
The uniform spin susceptibility calculated from SE+V diagrams 
in Fig.~\ref{fig:FeynmanDiagramFI}(c-e) is not vanishing, which reads
\begin{equation}
\chi_{\rm uni}(q_0) = I_{\rm SE}(q_0) + I_{\rm V}(q_0) \ne 0.
\label{chi_SE_V}
\end{equation}
It is pointed out that\cite{Chubukov2014} the seemingly higher order AL diagrams in Fig.~\ref{fig:FeynmanDiagramFI}(f-g)
are crucial in restoring the spin conservation near the magnetic QCPs.
The dynamical fermion-fermion interaction is mediated 
by the critical fluctuations in the spin channel.
The Stoner criterion reduces AL diagrams to 
the same order as self-energy and vertex diagrams
which restores the spin conservation with 
$\chi_{\rm uni}(q_0) = I_{\rm SE}(q_0) + I_{\rm V}(q_0) + I_{\rm AL}(q_0) = 0$.

In sharp contrast, the spin pumping setup in Fig.~\ref{fig:illustration}(a)
is at the magnetic heterostructure interface with two sources of magnetism,
namely, the itinerant electron spins in NFL metal and the local moments in FI.
It is important to note that the NFL is formed at an Ising nematic QCP in the charge channel.
And, the effective interaction between these non-quasiparticles is 
in the spin channel, which is mediated by the magnon excitations in the adjacent FI. 
The magnon has its own dynamics due to magnon-magnon interaction
as described in Eq.(\ref{FI_G}), which is fundamentally different from the Landau damping
inherited from the gapless fermions.
In this case, the AL diagrams is officially a higher order perturbation compared with SE+V diagrams.
As a result, the overall dynamic spin susceptibility at leading order, as given in Eq.(\ref{chi_SE_V}), is non-vanishing,
consistent with the expected breaking of spin conservation for the NFL metal.

To be concrete, we adopt the zero-temperature NFL Green function and 
calculate the uniform spin susceptibility by evaluating the SE and V diagrams in Fig.~\ref{fig:FeynmanDiagramFI}(c-d) and Fig.~\ref{fig:FeynmanDiagramFI}(e), respectively.
They are given by
\begin{equation}
\begin{aligned}
	I_{\rm SE}(q_0)=&6\int_k g_s^2(\bm{k},k_0)g_s(\bm{k},k_0-q_0) \Sigma_{\rm ex}(\bm{k},k_0) +(q_0\rightarrow -q_0),\\
	I_{\rm V}(q_0)=&2 \int_k \int_q \Gamma(\bm{q},k_0,q_0) g_s(\bm{k},k_0)g_s(\bm{k},k_0-q_0) ,\\
\end{aligned}
\label{SE+V}
\end{equation}
where $\int_k \equiv (2\pi)^{-3} \int dk_0d^2\bm{k}$ and $\int_q \equiv (2\pi)^{-3}\int dq_0^\prime d^2\bm{q}$.
$\Sigma_{\rm ex}$ and $\Gamma$ are the self-energy and vertex functions due to interfacial spin exchange coupling, respectively.
We define a modified interfacial coupling strength $\tilde{J} = J_1(16\pi v_{\rm F} N\sqrt{D})^{-1/2}$
and all the calculations below are provided up to leading order in $\mathcal{O}(\tilde{J}^2)$.
Moreover, to proceed analytically, we firstly focus on the the low frequency limit $|q_0|\ll \xi_{\rm b}^{-2} \equiv -\gamma_{\rm g} H$.
At the end of the section, we present the results for $|q_0|\sim \xi_{\rm b}^{-2} \equiv -\gamma_{\rm g} H$
and provide the detailed derivations in Supplementary Note 3.

We start with the SE diagram in Eq.(\ref{SE+V})
by evaluating the self-energy function $\Sigma_{\rm ex}$.
In the FM phase with a short correlation length $\xi_{\rm b} \ll 1$,
the self-energy function is given by
\begin{equation}
\begin{aligned}
	\Sigma_{\rm ex}(k_0)=&-\frac{J_1^2}{N}\int_q g_s(\bm{k}+\bm{q},k_0+q_0)G_0(\bm{q},q_0)\\
	\simeq & \tilde{J}^2 \big(\xi^{-1}_{\rm b} - 4ik_0 \xi_{\rm b} \big)+\mathcal{O}(\xi_{\rm b}).
	\end{aligned}
	\label{Sig_ex}
\end{equation}
In the low frequency limit $|q_0|\ll \xi_{\rm b}^{-2}$, we approximate the 
bare magnon Green function as $G_0^{-1}(\bm{q},q_0)\simeq iq_0-Dq_y^2-\xi_{\rm b}^{-2}-\alpha |q_0|$.
Substituting this expression into Eq.(\ref{SE+V}) and carrying out the ${\bm k}$-integral, we obtain
\begin{equation}
\begin{aligned}
I_{\rm SE}(q_0)=&\frac{-3J_1^2\xi_{\rm b}}{2\pi^2 Nv_{\rm F}^2\sqrt{D}}\int_0^{|q_0|}  \frac{dk_0 \ k_0}{\big[|q_0|-i\Delta \Sigma_{\rm F}(k_0,|q_0|)\big]^2}\\
	= &\frac{-3}{(1-a)} \frac{\tilde{J}^2 \xi_{\rm b}}{\pi v_{\rm F}c_{\rm F}^2 d^4_{{\bm k}_{{\rm F},s}}} |q_0|^{2-2a}+ \mathcal{O}(|q_0|^{2/3}),
\end{aligned}
\label{I_SE}
\end{equation}
where the functional difference is defined as $\Delta {\cal M}(x,y)\equiv {\cal M}(x-y) -{\cal M}(x)$,
and $a$  is the power of the frequency dependence in the renormalized fermionic Green function, which has been previously defined.
Next, we evaluate the V diagram in Eq. (\ref{SE+V}) starting from the vertex function
\begin{equation}
	\begin{aligned}
		&\Gamma(\bm{q},k_0,q_0)\\
		\equiv  &\frac{J_1^2}{N}\int_q g_s(\bm{k}+{\bm q},k_0+ q_0^\prime)g_s(\bm{k}+{\bm q},k_0+ q_0^\prime - q_0) G_0({\bm q},q_0^\prime)\\
=&-2\tilde{J}^2\int dq_0^\prime \frac{1}{\sqrt{\xi_{\rm b}^{-2}+\alpha|q_0^\prime|-iq_0^\prime}} \frac{\Delta\mathrm{sgn}(k_0+q_0',q_0)}{q_0-i\Delta \Sigma_{\rm F}(k_0+ q_0^\prime,q_0)} \\
\simeq &\frac{-2\tilde{J}^2\xi_{\rm b}}{(1-a)c_{\rm F} d^2_{\bm{k}_{{\rm F},s}}}|q_0|^{1-a}+\mathcal{O}(\xi_{\rm b}).
	\end{aligned}
\end{equation}
Similarly, by substituting this expression into Eq.(\ref{SE+V}) and carrying out the ${\bm k}$-integral, we obtain
\begin{equation}
	\begin{aligned}
		I_{\rm V}(q_0)=&\frac{-\tilde{J}^2 \xi_{\rm b}|q_0|^{1-a}}{(1-a)\pi v_{\rm F}c_{\rm F} d^2_{\bm{k}_{{\rm F},s}}}\int dk_0\frac{\mathrm{sgn}(k_0)\Theta[-k_0(k_0-q_0)]}{q_0-i\Delta \Sigma_{\rm F}(k_0,q_0)}\\
		\simeq &\frac{-\tilde{J}^2 \xi_{\rm b}}{(1-a)^2\pi v_{\rm F}c_{\rm F}^2 d^4_{\bm{k}_{{\rm F},s}}}|q_0|^{2-2a} + \mathcal{O}(|q_0|^{2/3}).\\
	\end{aligned}
	\label{I_V}
\end{equation}
From Eq.(\ref{I_SE},\ref{I_V}), we conclude that the contributions from SE and V diagrams 
are proportional to each other at the QCP in the presence of the NFL self-energy,
which leads to
\begin{equation}
	\begin{aligned}
	& I_{\rm V}(q_0) =I_{\rm SE}(q_0)/[3(1-a)],\\
	& \chi_{\rm uni}(q_0) = \frac{4-3a}{3-3a} I_{\rm SE}(q_0).\\
	\end{aligned}
	\label{prop_SE_V}
\end{equation}
We remark that (i) the generic scaling exponent of the uniform spin susceptibility in Eq.(\ref{scale_0})
can be explicitly expressed as $d_\chi = 2(z-d)=2$; 
(ii) This proportionality has been justified previously in the Fermi liquid phase 
with well-defined quasiparticles\cite{Chubukov2014}.

\begin{figure}[tb]
	\centering
	\includegraphics[width=0.9\linewidth]{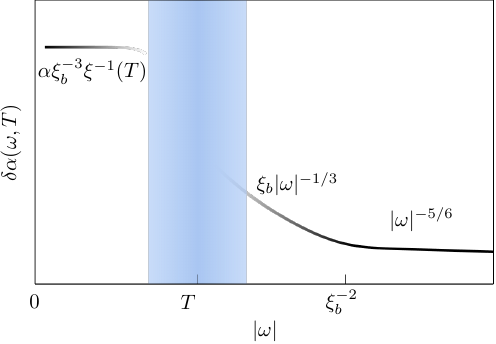}
	\caption{
	Schematic illustrations of the feromagnetic resonance modulations scaling forms.
	The frequency and temperature dependence of the enhanced Gilbert damping coefficient $|\delta \alpha(\omega,T)|$
	is plotted in distinct regimes.
	The low-frequency($|\omega|<T$, left) and high-frequency($|\omega|>T$, right) regimes
	are separated by a blue shaded region.
	}
	\label{fig:FMR}
\end{figure}

After analytic continuing to real frequency $ik_0 \rightarrow \omega + i0^+$, 
and evoking the correspondence in Eq.(\ref{Sigma_mag}),
we arrive at the retarded magnon self-energy correction in the regime $|\omega|\ll \xi_{\rm b}^{-2}$,
which reads
\begin{equation}\label{Sigma_m_0}
    \Pi_{{\bm k}=0}(\omega) = \frac{-9\tilde{J}^2\xi_{\rm b}}{\pi v_{\rm F}c_{\rm F}^2 d^4_{{\bm k}_{{\rm F},s}}}|\omega|^{2/3} \Big[1 - \sqrt{3}i{\rm sgn}(\omega) \Big].
\end{equation}
Accordingly, the FMR modulations, namely the resonance frequency shift and the enhanced Gilbert damping, 
acquire peculiar scalings in frequency
\begin{equation}\label{FMR_mod0}
    \delta \alpha \sim \xi_{\rm b} {\rm sgn}(\omega) |\omega|^{-\frac{1}{3}}, \quad \delta H \sim \frac{\xi_{\rm b}}{\gamma_{\rm g}} {\rm sgn}(\omega)|\omega|^{2/3}.
\end{equation}
These power scalings are schematically plotted in Fig.~\ref{fig:FMR}
and are subjected to experimental validations.
In the low energy and zero temperature limit: $T<|\omega|< \omega_{\rm c}$,
the diverging coefficient $\delta \alpha$ indicates a different spin relaxation mechanism, which is in sharp contrast to the conventional linear-in-$\omega$ Gilbert damping.
The exponents reflect the universal scaling behavior near the Ising nematic QCP
in the dynamic limit.
Near the resonant frequency $|\omega|\sim \xi_{\rm b}^{-2}$, we show that [see Supplementary Note 3a] one can simply 
replace $\xi_{\rm b}$ in Eq.~(\ref{FMR_mod0}), which leads to $\delta \alpha \sim {\rm sgn}(\omega) |\omega|^{-5/6}$.
The FMR modulations take distinct set of scaling forms where
the enhanced Gilbert damping coefficient becomes even more divergent.
Finally, we point out that by comparing with the quasiparticle weights,
the FMR modulations capture the disappearance of the quasiparticles in the low-energy limit.

\smallskip
\noindent\textbf{FMR modulations at finite temperatures}.
\label{subsec_4}
The finite-temperature properties of NFLs in the quantum critical regime mediated by the Ising nematic QCP 
cannot be simply inferred from the zero-temperature results.
The quantum fluctuation of the bosons obeys the $\omega/T$ scaling rule
and leads to $T^{\frac{2}{3}}$ dependence for fermion self-energy;
while, the contribution from thermal fluctuation is drastically different and dominates at low temperatures~\cite{Metzner2006}.
Importantly, it is demonstrated that vertex correction is subdominant 
when both thermal and quantum fluctuations are included~\cite{Punk2016}.
As a result, one can write down the finite temperature version of Eliashberg equations
which leads to the solution
\begin{subequations}\label{eq:ge}
    \begin{align}
        g^{-1}_{\rm R}(\mbf{k},\omega) & = \omega+i0^+ - \epsilon_{\bm k} + i  \gamma_{{\bm k}_{\rm F}}(T),\\
        D_{\rm R}^{-1}(\mbf{q},\Omega) & = \xi^{-2}(T) + a|{\bm q}|^2 -i b\frac{\Omega}{\gamma(T)}. 
    \end{align}
\end{subequations}
where $a$ and $b$ are nonuniversal constants. 
The boson correlation length and electron scattering rate scale as 
$\xi^{-1}(T)\sim \sqrt{T\ln(\epsilon_{\rm F}/T)}$ and $\gamma(T)\sim \sqrt{T/\ln (\epsilon_{\rm F}/T)}$, respectively. 
We note that the expressions in Eq.(\ref{eq:ge}) can not be deduced from the opposite $T=0$ limit
by applying the $\omega/T$ scaling. This is due to the $T$-dependence of the boson correlation
arising from the dangerously irrelevant boson interactions.
The imaginary part of the retarded fermion self-energy at $T\gg |\omega|$
is derived as $\gamma_{{\bm k}_{\rm F}}(T) = \lambda^2 d^2_{{\bm k}_{\rm F}} T\xi(T) / (4v_{\rm F}\sqrt{a})$ for $T\ll T_{0}$. 
The upper energy bound $T_0= \epsilon_{\rm F} e^{-\lambda^2/v_{\rm F}^2}$ is specified by the condition 
$|\gamma_{{\bm k}_{\rm F}}(T)| \ll \xi^{-1}(T) v_{\rm F}/\sqrt{a}$ within which the solutions in Eq.(\ref{eq:ge}) hold.
Moreover, we focus on the quantum critical regime $|\omega| \ll |\gamma_{{\bm k}_{\rm F}}(T)|$
where the fermion self-energy overwrites the bare dynamics.
The finite-$T$ expression in Eq.(\ref{eq:ge}a) can be further simplified as
$\mathrm{Im}g_{\rm R}(\bm{k},\omega)\simeq \gamma_{\bm{k}_{\rm F}}(T)/\big[(v_{\rm F} k_x)^2+\gamma_{\bm{k}_{\rm F}}^2(T)\big]$.

In parallel with the previous section, we evaluate the uniform spin susceptibility
in the quantum-critical finite-temperature fan from the SE+V diagrams.
The calculation details are written in Supplementary Note 4 and 5.
Similarly, we first calculate the fermion self-energy due to interfacial spin exchange interaction
where the local moments in FL are in the FM ordered phase with small correlation length $\xi^{-2}_{\rm b} \gg T(\gg |\omega|)$.
The imaginary part of retarded component is given by
\begin{equation}
	\begin{aligned}
	\mathrm{Im}[\Sigma_{\rm ex}^{\rm R}(|\omega| \ll T)] &= -\frac{J_1^2}{4\pi^3 N}\int d^2\bm{q} d\Omega [n_{\rm B}(\Omega)+n_{\rm F}(\Omega)]\\
	&\quad \quad \quad  \times \mathrm{Im}g_{\rm R}(\bm{q}+\bm{k},\Omega)\mathrm{Im}G_0(\bm{q},\Omega) \\
		  \simeq& -4 \alpha\tilde{J}^2   \xi^3_{\rm b} T^2+\mathcal{O}(\xi_{\rm b}^3). \\
	\end{aligned}
\end{equation}
Using the Lehmann representation, we obtain the expression at Matsubara frequencies as 
\begin{equation}
\begin{aligned}
&\Sigma_{\rm ex}(ik_0,T)
=  \int_{-\omega_{\rm c}}^{+\omega_{\rm c}} \frac{d\omega}{\pi}\frac{\mathrm{Im}\Sigma_{\rm ex}^{\rm R}(\omega,T)}{\omega-ik_0} \\
& \simeq  \int_{-T}^{+T} \frac{d\omega}{\pi}\frac{\mathrm{Im}\Sigma_{\rm ex}^{\rm R}(|\omega|\ll T)}{\omega-ik_0} 
=-i\mathrm{sgn}(k_0)4\alpha \tilde{J}^2 \xi^3_{\rm b} T^2.
\end{aligned}
\end{equation}
We note that the dominant contribution in the frequency integral
comes from the finite-temperature regime approximately bounded by $|\omega|\le T$.
Then, we substitute this expression into $I_{\rm SE}(q_0,T)$
and sum over the Matsubara frequencies,
which yields
\begin{equation}
	\begin{aligned}
	I_{\rm SE}(q_0,T)
	=&-\frac{24\alpha \tilde{J}^2 }{\pi v_{\rm F}}\xi_{\rm b}^3 T^2\frac{|q_0|}{\big[|q_0| +2\gamma_{\bm{k}_{\rm F}}(T)\big]^2}\\
	\simeq & -\frac{6\alpha \tilde{J}^2 \xi_{\rm b}^3 }{\pi v_{\rm F}}\frac{T^2}{{\gamma_{\bm{k}_{\rm F}}}^2(T)}|q_0|.\\
	\end{aligned}
	\label{I_SE_T}
\end{equation}

For the V diagram, we first calculate the vertex function,
\begin{equation}
	\begin{aligned}
	&\Gamma(\bm{q},k_0,q_0;T)\\
	=&-\frac{\alpha \tilde{J}^2\xi_{\rm b}^3}{\gamma_{\bm{k}_{\rm F}}(T)} \int  d\Omega \Omega\Big\{-2\mathrm{sgn}(k_0)\mathrm{sgn}(k_0-q_0)n_{\rm B}(\Omega) \\
	& +\big[\mathrm{sgn}(q_0)\mathrm{sgn}(q_0-k_0)+\mathrm{sgn}(q_0)\mathrm{sgn}(k_0)\big]n_{\rm F}(\Omega)
	 \Big\}\\
	\end{aligned}
	\label{Vertex}
\end{equation}
where the bare magnon self-energy  at finite-$T$ is approximated as $\mathrm{Im}G_0(\bm{q},\Omega)=-\alpha\Omega/[(Dq_\parallel^2+\xi_{\rm b}^{-2}-\Omega)^2+\alpha^2\Omega^2]\simeq -\alpha\Omega/(Dq_\parallel^2+\xi_{\rm b}^{-2})^2$.
Substituting this expression into Eq.(\ref{SE+V}) and carrying out the ${\bm k}$-integral, we obtain:
\begin{equation}
	\begin{aligned}
		I_{\rm V}(q_0,T)=&-\frac{\alpha\tilde{J}^2\xi_{\rm b}^3}{\pi v_{\rm F} \gamma_{\bm{k}_{\rm F}}(T)} \int  d\Omega\frac{|q_0|2\Omega [n_{\rm B}(\Omega)+n_{\rm F}(\Omega)]}{|q_0|+2\gamma_{\bm{k}_{\rm F}}(T)}\\
\simeq &-\frac{2\alpha\tilde{J}^2}{\pi v_{\rm F} \xi_{\rm b}^3} \frac{T^2}{{\gamma_{\bm{k}_{\rm F}}}^2(T)} |q_0|.\\
	\end{aligned}
	\label{I_V_T}
\end{equation}
By comparing Eq.(\ref{I_SE_T},\ref{I_V_T}), we conclude that the proportionality between SE and V 
in Eq.(\ref{prop_SE_V}) continuities to hold at finite temperatures 
in the quantum critical regime $|q_0| \ll |\gamma_{{\bm k}_{\rm F}}(T)|$.

Finally, we arrive at the retarded finite-temperature magnon self-energy 
at leading order in $\mathcal{O}(\tilde{J}^2)$, which reads
\begin{equation}
	\begin{aligned}
	\Pi_{{\bm k}=0}(\omega,T)\sim \frac{v_{\rm F} a \tilde{J}^2}{\pi\lambda^2 d^2_{{\bm k}_{\rm F}} \xi_{\rm b}^{3}}\xi^{-2}(T) (i\alpha\omega),
	\end{aligned}
	\label{Sig_mg_T}
\end{equation}
where the real part is at the next order taking a form 
$\sim - (\omega^2 T^2)/\gamma_{{\bm k}_{\rm F}}^3(T)$ and is therefore omitted.
This expression is consistent with the generic scaling form in Eq.\eqref{scale_1} 
with $d^\prime_{\chi} =-2$ at finite temperatures.
Accordingly, the enhanced Gilbert damping coefficient is given by
\begin{equation}\label{FMR_mod_T}
    \delta \alpha \sim \alpha \xi^{-2}(T),
\end{equation}
and the resonant frequency shift vanishes $\delta H \simeq 0$ in the quantum critical fan region,
which are schematically plotted in Fig.~\ref{fig:FMR}.
We note that, in the quantum critical fan at finite-$T$, the Gilbert damping mechanism is restored,
in contrast to the divergent Gilbert damping coefficient at $T=0$ in Eq.(\ref{FMR_mod0}).
The information of the correlation length can be extracted from the FMR modulations which, 
at the same time, reflects the non-quasiparticle nature of the NFL metal and the underlying QCP.
To see this, we evaluate the real part of the fermionic self-energy 
$\mr{Re}\Sigma_{\rm F}(\mbf{k}_{\rm F},\omega)\sim -d^2_{\mbf{k}_{\rm F}} \omega\xi(T)$ for $|\omega|<T<\omega_{\rm c}$ which leads to a vanishing quasiparticle weight due to the divergent correlation length [see Supplementary Note 6]
\begin{equation}
    Z(\mbf{k}_{\rm F},\omega;T)\simeq d_{\mbf{k}_{\rm F}}^{-2}\xi^{-1}(T),\quad k_{{\rm F},x}\neq \pm k_{{\rm F},y}.
\end{equation}

\bigskip
\noindent\textbf{\large DISCUSSION}\\
We have presented a practical implementation of the magnetic heterostructure
that probes the many-body spin/charge correlations.
We study the FMR-driven spin pumping effect at the interface of NFL/FI heterostructure in the absence of quasiparticles.
The NFL metal is induced near a Pomeranchuk-type of QCP in the charge channel;
while, the dynamic spin correlation function of the NFL metal is non-vanishing
and acquires universal power-law scalings in frequency and temperature domains,
due to its interfacial exchange coupling to the adjacent FI.
The experimental measurable FMR modulations, namely the resonance frequency shift and the enhanced Gilbert damping, convey valuable information characterizing the NFL behaviors and the disappearance of the quasiparticles. 
We conclude that the spintronics experiments, particularly spin pumping, 
can take full advantage of the magnetic heterostructure, meanwhile, 
shed light on the non-quasiparticle feature of spin relaxation in NFLs.
Our proposal is also helpful in reconciling the current theoretical debates in
the scaling forms of dynamic correlation functions near Ising nematic QCP.

The magnetic heterostructure is far from simple stacking of two different materials.
The heterostructure by design breaks inversion symmetry,
thus, allowing the existence of anti-symmetric spin exchange interaction,
which is also known as the Dzyaloshinski--Moriya interaction (DMI).
On one hand, the DMI can drive the magnetic insulators into forming more complex magnetic orders,
such as the skyrmion lattice structures where local magnetic moments swirl spatially 
in a non-collinear way while forming a periodical lattice.
The formation of more complex magnetic ordered states inevitably 
makes the phenomena of the spin pumping effect more diverse and interesting.
It has already been reported that the magnetic fluctuation of the skyrmion lattices 
can mediate exotic type of electron interaction in normal metals.
Owing to the non-collinear nature of the magnetic ground state, 
topological superconductivity can be induced directly~\cite{Sudbo2023}.
The investigation of the competiting role between 2D topological superconductivity and novel type of NFLs is an intriguing topic. Moreover, the heterostructure setup offers an ideal platform for studying the spin pumping effect, which is particularly appealing to the spintronics community.
On the other hand, the DMI is an essential ingredient when the magnetic insulator approaches its criticality.
Infinite many critical bosons can simultaneously reach criticality due to the presence of DMI
in three-dimensional chiral magnet~\cite{Zhang2023} as well as 2D magnetic heterostructure interface~\cite{Zhang2023B}.
The infinite many critical boson can induce a different type of NFL state for the itinerant magnets nearby~\cite{Lake2021,YBKim2022,Raghu2023,Zhang2023,Zhang2023B}.

Finally, we remark on the non-equilibrium theoretical framework adopted in the present study.
The FMR modulations are derived perturbatively in terms of 
interfacial exchange coupling strength $J_1$ in the Keldysh representation.
In contrast, the self-energy correction of the NFL is independent of $J_1$,
that is, the stability of the NFL state is assumed as prior.
From a theoretical point of view, it is greatly desired that the coupled 
two sides of the magnetic heterostructure are solved self-consistently; yet, it poses great difficulty.
Thanks to the development of Sachdev-Ye-Kitaev (SYK) model~\cite{Chowdhury2022},
the magnetic exchange interaction can be regarded as a random variable in fictitious flavor space.
The replica trick used in solving SYK model can lead to 
a solvable equation set for the entire magnetic heterostructure.
The mutual interactions between the itinerant magnets and driven magnetic insulator
can be determined self-consistently via numerical calculations.
This greatly improves the current theoretical treatments in analyzing the spin pumping effect,
particularly, the non-steady non-equilibrium spin dynamics.

\bigskip
\noindent \textbf{\large Data Availability}\\
Data sharing not applicable to this article as no datasets were generated or analysed during the current study.

\bigskip

\def\bibsection{\ } 
\noindent \textbf{\large REFERENCES}
\bibliographystyle{naturemag}
\bibliography{Ref}

\bigskip

\noindent \textbf{\large ACKNOWLEDGEMENTS}\\
X.-T.Z. is supported by National Natural Science Foundation of China(NSFC) Grant No.~12404178.
X.-T.Z. acknowledges valuable discussions and financial support from Prof. Fu-Chun Zhang 
funded by China's Ministry of Science and Technology (Grant No.~2022YFA1403902) and NSFC (Grant No.~11920101005). 
L.Z. is supported by the National Key R\&D Program of China (Grant No.~2018YFA0305800), the NSFC (Grant No.~12174387), the Strategic Priority Research Program of CAS (Grant No.~XDB28000000), and the CAS Youth Innovation Promotion Association.
M.M. is supported by the National Natural Science Foundation of China (NSFC) under Grant No. 12374126, by the Priority Program of Chinese Academy of Sciences under Grant No. XDB28000000, and by JSPS KAKENHI for Grants (Nos. 21H01800, 21H04565, 23H01839, and 24H00322) from MEXT, Japan.

\bigskip

\noindent \textbf{\large AUTHOR CONTRIBUTIONS}\\
X.-T.Z. and M.M. conceived the original idea,
and Y.O. attended the discussions.
X.-T.Z. designed and supervised this project. 
X.-T.Z. and Y.-H.X. performed the theoretical calculation. 
X.-T.Z., Y.-H.X., X.-P.Y. and L.Z. wrote the manuscript.
X.-P.Y. and X.-T.Z. plotted the schematic figures.

\bigskip
\noindent \textbf{\large COMPETING INTERESTS}\\
The authors declare no competing interests.

\end{document}